\documentclass[twocolumn,times]{aastex631}

\usepackage{color}

\graphicspath{{./}{fig/}}

\begin{document}

\title{Discovery of Rapid Polarization Angle Variation During the 2022 Outburst of XTE J1701-462
}

\author[0000-0001-9893-8248]{Qing-Chang Zhao}
\affiliation{Key Laboratory of Particle Astrophysics, Institute of High Energy Physics, Chinese Academy of Sciences, Beijing 100049, China} 
\affiliation{University of Chinese Academy of Sciences, Chinese Academy of Sciences, Beijing 100049, China} 

\author{Hong Li}
\email{lihong02@ihep.ac.cn}
\affiliation{Key Laboratory of Particle Astrophysics, Institute of High Energy Physics, Chinese Academy of Sciences, Beijing 100049, China} 

\author[0000-0002-2705-4338]{Lian Tao}
\email{taolian@ihep.ac.cn}
\affiliation{Key Laboratory of Particle Astrophysics, Institute of High Energy Physics, Chinese Academy of Sciences, Beijing 100049, China} 

\author[0000-0001-7584-6236]{Hua Feng}
\email{hfeng@ihep.ac.cn}
\affiliation{Key Laboratory of Particle Astrophysics, Institute of High Energy Physics, Chinese Academy of Sciences, Beijing 100049, China} 

\begin{abstract}

The geometry of the Comptonization corona in neutron star low-mass X-ray binaries is still unclear. We conducted time-resolved polarimetric analysis of the archival observations of XTE J1701--462 obtained with the \textit{Imaging X-ray Polarimeter Explorer} during its 2022 outburst, and found that the polarization angle (PA) varied significantly with time when the source was in the normal branch (NB), with $67 \pm 8^{\circ}$ in the first epoch, $-34 \pm 8^{\circ}$ in the second, and $-58 \pm 8^{\circ}$ in the third, last epoch. Meanwhile, the polarization degree remained constant at around 2\%, above the minimum detectable polarization at the 99\% confidence level (MDP$_{99}$). The rapid PA variation causes depolarization in the time-averaged data, resulting in a nondetection as reported in the literature. The rapid (intra-day) PA variation may suggest that there is a fast transformation of the corona geometry, likely switching from a slab geometry with enhanced disk emission and reflection, to a more vertically extended spreading layer geometry. 

\end{abstract}


\section{Introduction}
\label{sec:intro}

Neutron star low mass X-ray binaries (NS-LMXBs) are powered by weakly magnetized (typically $10^{7}-10^{9}$~G) neutron stars accreting matter from a companion star. A widely used classification scheme for NS-LMXBs is based on their evolutionary tracks in the color-color diagrams (CCD) and/or the hardness-intensity diagrams \citep[HID;][]{Hasinger_LMXB_CCD, van_der_klis_ccd}. According to the shape of evolutionary tracks, the sources can be classified into two subclasses: ‘Z’ sources or ‘Atoll’ sources.

The X-ray emission from NS-LMXBs generally consists of two spectral components: a soft thermal component and a hard Comptonization component. The soft component can be modeled as a multicolor disk blackbody with an innermost disk temperature of $\lesssim 1$ keV. The hard Comptonization component is produced in a relatively cool corona (2--3 keV) likely located in the transition layer (or boundary layer), a region between the Keplerian accretion disk and the surface of neutron star \citep{Shakura_boundary_layer, Popham_boundary_layer}. Frequency-resolved X-ray spectroscopy suggests that the short-term X-ray variability mainly originates in this region \citep{Gilfanov_2003}. The geometry of the transition layer is unclear \citep[for reviews see][]{Done_2007, Di_Salvo_LMXB}. For example, it is proposed that a spreading layer over the neutron star surface will form when the accretion rate is high \citep{Inogamov_SL, Suleimanov_SL}. X-ray spectro-timing-polarimetry, on the other hand, has the potential to break the degeneracy in a model constrained by a single technique, and help resolve the geometry of the transition layer \citep[e.g.,][]{Farinelli_2024}. In particular, the polarization angle (PA) provides insightful information about the underlying geometry. This has been successfully applied onto several systems. 

Sco X-1 is the first NS-LMXB observed with X-ray polarimetry; both OSO-8 \citep{Novick1977} and PolarLight \citep{Polarlight} found a PA in line with the orientation of radio jet \citep{Long1979,Long_scox-1}, which is presumably the symmetric axis of the inner accretion flow.
The \textit{Imaging X-ray Polarimetry Explorer} \citep[\textit{IXPE};][]{Soffitta_ixpe, Weisskopf_ixpe} found a similar result in Cyg X-2 that the PA is aligned with the radio jet \citep{Farinelli_cygx-2}. Such an alignment may suggest that the X-ray emission arises from the spreading layer. However, the \textit{IXPE} observation of Sco X-1 detected a distinct PA,
suggestive of a variable emitting geometry probably related to the emission state \citep{La_monaca_scox-1}.
In Cir X-1, the PA was found to vary with both time and hardness ratio, by $67^{\circ} \pm 11^{\circ}$ between the states with the lowest and highest hardness ratios \citep{Rankin_cirx-1}, indicative of a synchronous change in emitting geometry and spectrum.
Notably, GX 13+1 exhibited a continuous rotation of PA by $70^{\circ}$ over a course of two days, along with a significant variation in the polarization degree (PD); however, no significant spectral changes were identified \citep{Bobrikova_GX_13+1}. 
A PD variation was observed in both GX 5-1 \citep{Fabiani_GX5-1} and XTE J1701--462 \citep{Cocchi_xtej1701, Jayasurya_xtej1701, Yu_xtej1701} while the PA remained constant; also see  \citet{Ursini_etal_2024_review} for a recent review. 

XTE J1701--462 is a transient NS-LMXBs that was first discovered with the All-Sky Monitor (ASM) onboard the \textit{Rossi X-ray Timing Explorer} (\textit{RXTE}) on January 18, 2006 \citep{Remillard_xtej1701}. This source is particularly intriguing, as it experienced all the known spectral states of the NS-LMXB during the 2006 outburst \citep{Homan_xtej1701, Lin_xtej1701}. The distance to XTE J1701--462 is estimated to be 8.8 kpc with an uncertainty of 15\%, based on the photospheric radius expansion of type-I bursts \citep{Lin_xtej1701_distance}. Furthermore, the absence of eclipses or absorption dips constrains the orbital inclination to be less than $75^{\circ}$ \citep{Lin_xtej1701}.

XTE J1701--462 underwent a new outburst in September 2022 \citep{Iwakiri_1701_new_outbust}, during which the source was observed with \textit{IXPE} \citep{Cocchi_xtej1701, Jayasurya_xtej1701}, with a high PD of approximately 4.6\% in the horizontal branch (HB). However, the polarization was undetected when the source entered the normal branch (NB), with a 99\% PD upper limit of 1.5\%. 
\citet{Yu_xtej1701} reanalyzed the data and obtained consistent results.

In this work, we analyzed the archival \textit{IXPE} data of XTE J1701--462 and discovered a significant PA variation in the second observation when the source was in the NB. The paper is organized as follows. Details about the observations and data reduction are described in Section \ref{sec:data_reduction}. The results are presented in Section \ref{sec:results} and discussed in Section \ref{sec:disc_conclu}.

\section{Observations and Data Reduction}
\label{sec:data_reduction}

\begin{figure}[tb]
\centering
\includegraphics[width=0.4\textwidth]{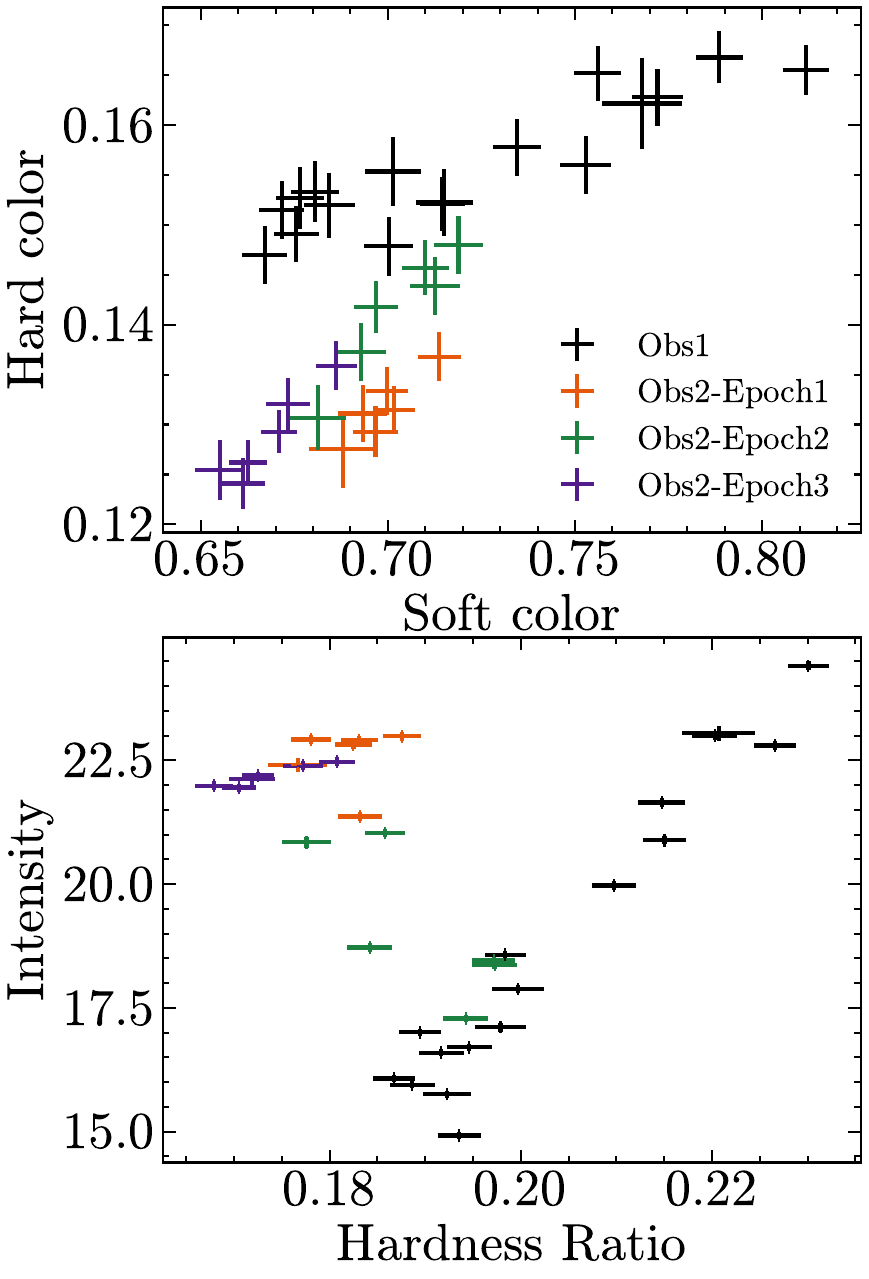}
\caption{CCD (top) and HID (bottom) of XTE J1701--462 constructed using the IXPE data. The color or hardness is defined as the ratio of count rate in two bands: (3--5 keV) / (2--3 keV) for the soft color, (5--8 keV) / (3--5 keV) for the hard color, and (4--8 keV) / (2--4 keV) for the hardness ratio. The intensity is the count rate in 2--8 keV.
}
\label{fig:hid_ccd}
\end{figure}

The source was observed with \textit{IXPE} in the HB on 2022-09-29 (ObsID 01250601; Obs1) and in the NB on 2022-10-08 (ObsID 01250701; Obs2), see Figure~\ref{fig:hid_ccd} for the CCD and HID in the two observations.

Our analysis begins with the level-2 data, which are further reduced and analyzed using \texttt{ixpeobssim v31.0.1} \citep{Baldini_obssim} and \texttt{HEAsoft v6.33.2}. 
We selected a circular region with a radius of 90$\arcsec$ for source extraction. Background subtraction was not performed as suggested due to the high count rate of the source \citep{DiMarco_background}.

The source events in the 2--8 keV energy range are selected using \texttt{xpselect}. To conduct a model-independent polarimetric analysis, polarization cubes are generated using the \texttt{PCUBE} algorithm in \textsc{ixpeobssim}. Furthermore, we produced the Stokes parameters spectra --- \textit{I}, \textit{Q}, and \textit{U} --- using the \texttt{PHA1}, \texttt{PHA1Q}, and \texttt{PHA1U} algorithms, respectively. The Stokes \textit{I} spectra are grouped to ensure a minimum of 30 counts per bin, while a constant energy binning with a bin size of 0.2 keV is applied to the Stokes \textit{Q} and \textit{U} spectra to facilitate spectro-polarimetric analysis in \textsc{xspec} \citep{Arnaud_xspec}.

The Nuclear Spectroscopic Telescope Array (\textit{NuSTAR}) \citep{Harrison_nustar} conducted an observation of XTE J1701--462 that partially overlaps in time with \textit{IXPE} Obs2. Cleaned level-2 events were extracted using the \texttt{nupipeline} routine from the \textit{NuSTAR} Data Analysis Software (\texttt{NuSTARDAS}) package. Source events were collected from a circular region with a radius of 60$\arcsec$, while background events were extracted from a concentric annular region from 120$\arcsec$ to 150$\arcsec$. Energy spectra in the 3--30~keV band, where the source dominates, were generated using \texttt{nuproducts}, and rebinned to have a minimum of 30 counts per bin.

\begin{figure}[tb]
\centering
\includegraphics[width=0.5\textwidth]{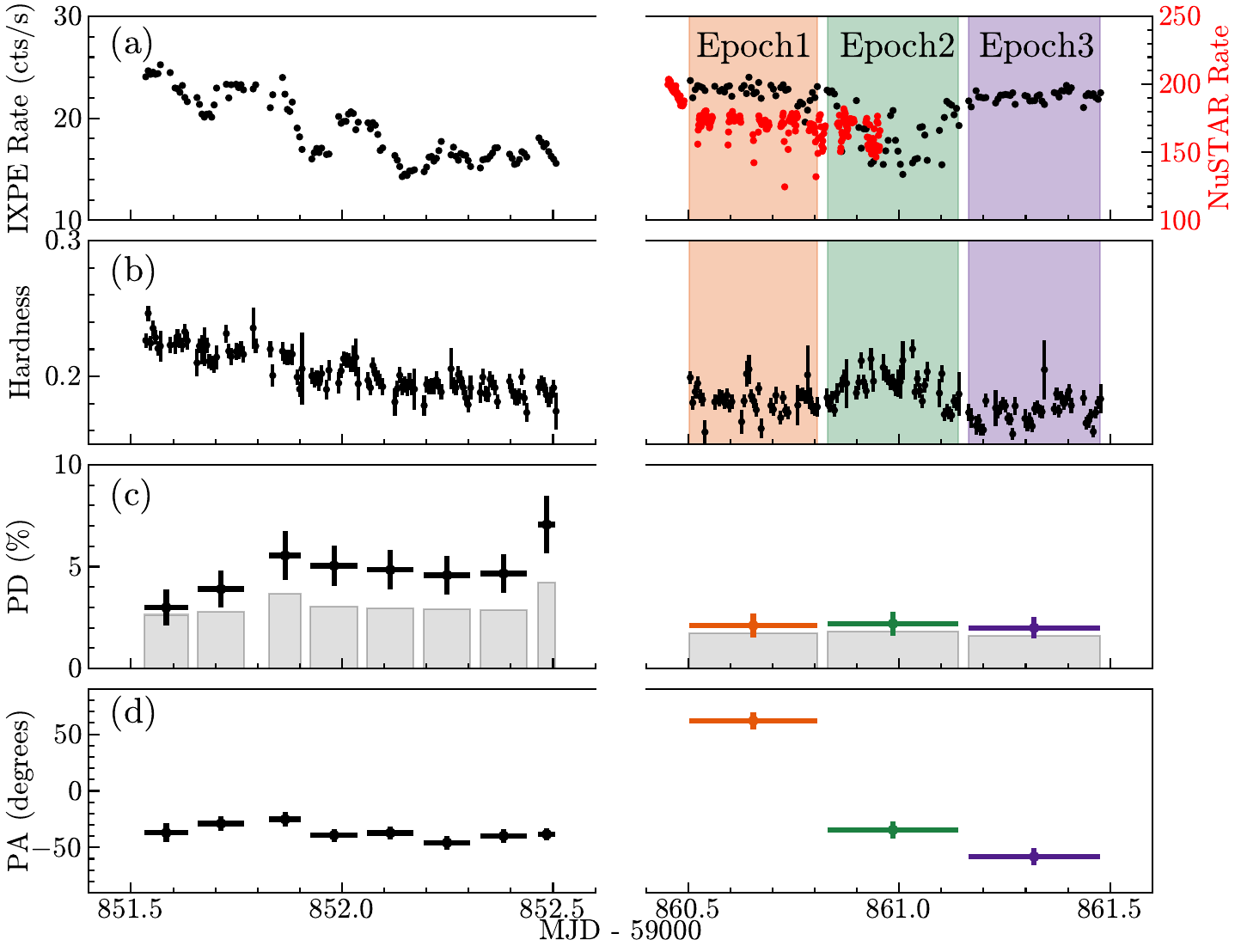}
\caption{Time variation of the spectral and polarization properties. \textbf{(a)}: \textit{IXPE} DU1 light curve in the 2--8 keV energy range (black) and \textit{NuSTAR} light curve in 3--30 keV (red). \textbf{(b)}: Hardness ratio as count rate in 4--8 keV to that in 2--4 keV. \textbf{(c)} and \textbf{(d)}: PD and PA in different time epochs. The gray bars mark the MDP$_{99}$. The three epochs in Obs2 are highlighted.}
\label{fig:time_resolved}
\end{figure}

\section{Analysis and Results}
\label{sec:results}

\subsection{Polarimetric Analysis}

First, we performed a model-independent polarimetric analysis using the \texttt{PCUBE} algorithm. We found time-averaged ${\rm PD} = 4.5\% \pm 0.4\%$ and ${\rm PA} = -37^{\circ} \pm 2^{\circ}$ in Obs1, and a nondetection in Obs2 (${\rm PD} = 0.84\% \pm 0.33\%$), consistent with previous results \citep{Cocchi_xtej1701, Jayasurya_xtej1701, Yu_xtej1701}.

We further conducted spectro-polarimetric analysis. Following the approach of \citet{Cocchi_xtej1701}, we modeled the disk emission with \texttt{diskbb} and the transition layer emission with \texttt{bbodyrad}.
To resolve cross-calibration discrepancies, we multiplied the model spectra by $KE^{\Delta \Gamma}$ \citep{Zdziarski_2021}, and fixed $K = 1$ and $\Delta \Gamma = 0$ for DU1. The \texttt{polconst} model is adopted to estimate the polarization. For Obs1, the PD is found to be $4.7\% \pm 0.3\%$ and the PA is $-36^{\circ} \pm 2^{\circ}$. For Obs2, the result is still a nondetection with ${\rm PD} = 0.6\% \pm 0.3\%$. The results are consistent with those obtained with \texttt{PCUBE} within errors.

\begin{figure}[tb]
\centering
\includegraphics[width=0.45\textwidth]{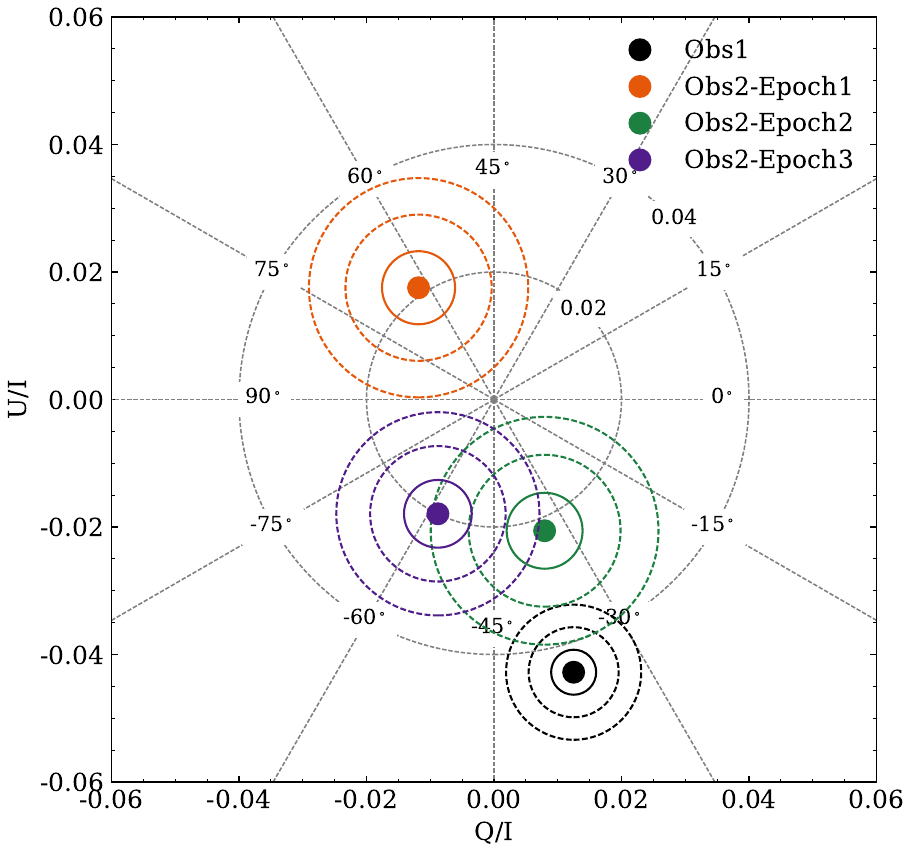}
\caption{Normalized Stokes parameters \(U/I\) and \(Q/I\) over the energy range of 2--8 keV. The contours represent confidence levels at 1\(\sigma\), 2\(\sigma\), and 3\(\sigma\). 
}
\label{fig:contour}
\end{figure}

We then performed a time-resolved analysis. In each observation, we grouped the data into several segments with the same number of satellite orbits in each and assured that PD $>$ MDP$_{99}$ (the minimum detectable polarization at the 99\% confidence level). 
For Obs1, where the PD is relatively high, every two orbits are grouped into a segment except the last orbit falling into a single one. 
For Obs2, every five orbits are grouped into a segment. 
The PD and PA variations as a function of time are shown in Figure~\ref{fig:time_resolved}.
In Obs1, both PD and PA are consistent with a constant over time.
However, in Obs2, a significant time variation of PA is observed, at $67^{\circ} \pm 8^{\circ}$ (Epoch1), $-34^{\circ} \pm 8^{\circ}$ (Epoch2), and $-58^{\circ} \pm 8^{\circ}$ (Epoch3), respectively in the three epochs, while the PD remained constant at about 2\%\footnote{We note that consistent results were briefly reported in \citet{DiMarco_2024_nslmxb} during the review process of this paper.}. The normalized Stokes parameters $U/I$ and $Q/I$ for Obs1 and the three epochs in Obs2 are displayed in Figure~\ref{fig:contour}.

\subsection{Spectral Analysis}

Simultaneous \textit{NuSTAR} and \textit{IXPE} data are used for spectral modeling in Epoch1 and Epoch2 (no \textit{NuSTAR} data in Epoch3), to investigate possible spectral variations associated with the observed PA variation. 
We began with the \texttt{Tbabs(Diskbb + Bbodyrad)} model. A weak reflection feature was identified in the residuals. Therefore, we included the \texttt{RelxillNS} model \citep{RelxillNS} to account for reflection. In this model, we fixed the density $\log (N / {\rm cm}^{-3}) = 19$, as the fitting result is insensitive to this parameter, the inner radius at \(R_{\rm ISCO}\) and the outer radius at \(1000\,R_{\rm g}\), and the radial profile of emissivity at $-3$. The dimensionless spin \( a \) is given by \( a = 0.47 / P_{\rm ms} \) \citep{Braje_NS}. For typical NS spin periods ranging from 1.5 to 5\,ms \citep{Patruno_NS_Spin}, \( a \approx 0.1\)–\(0.3\). Testing various values, \( a = 0.1 \) provided the best fit and was fixed in our spectral analysis.
The \( KE^{\Delta \Gamma} \) correction for cross-calibration issues is also included. 
The temperature \( kT_{\rm bb} \) of \texttt{RelxillNS} is not tied to \( kT \) of \texttt{Bbodyrad}, as coupling them significantly degrades the fit quality ($\Delta{\chi^2}=180$) and leads to an unphysical dominance of the disk flux ($\sim$ 90\% in 2--8 keV). The temperature discrepancy may arise because the observed boundary layer region differs from the region illuminating the disk.

\begin{figure*}[tb]
\centering
\includegraphics[width=0.32\textwidth]{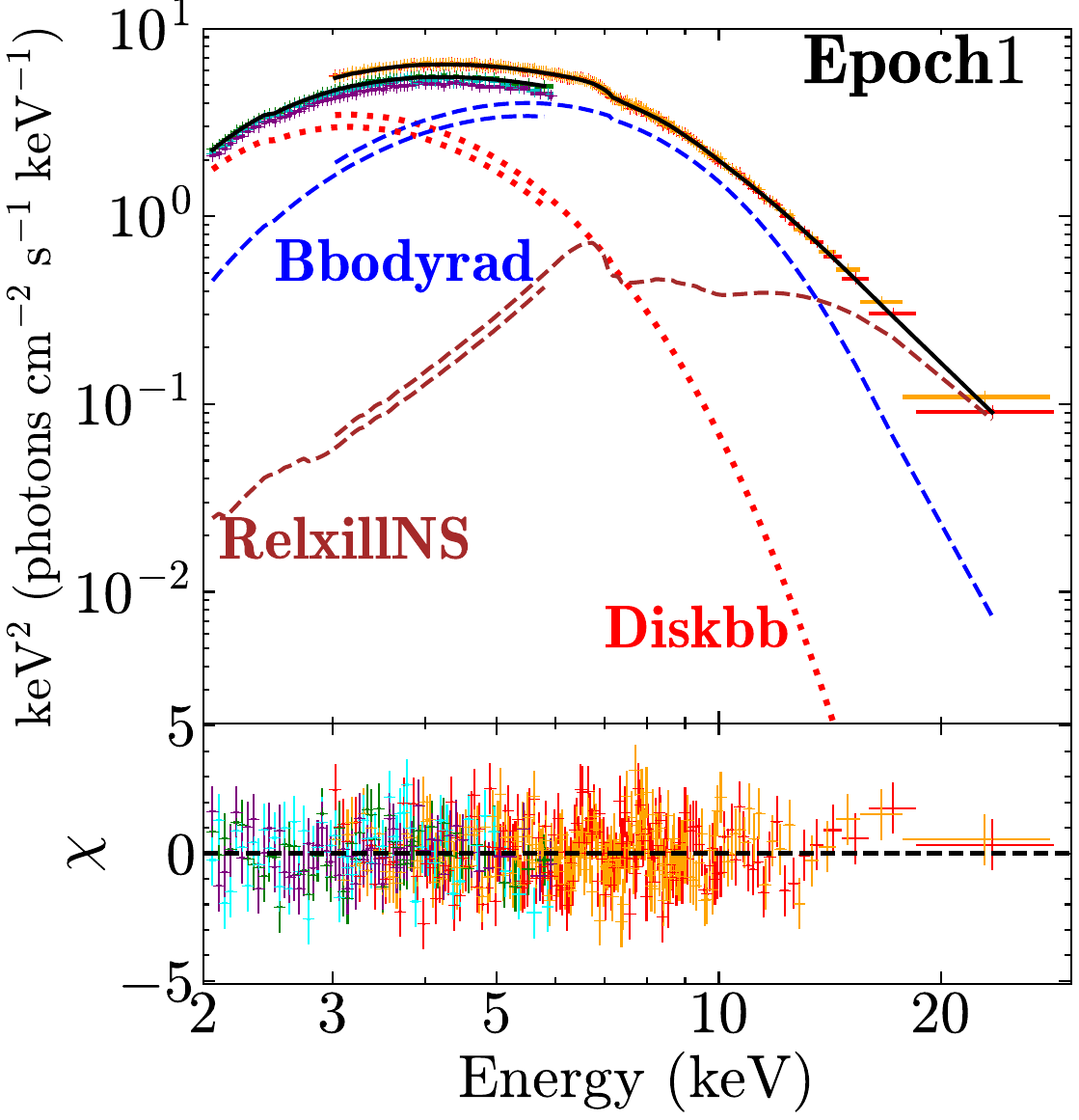}
\includegraphics[width=0.32\textwidth]{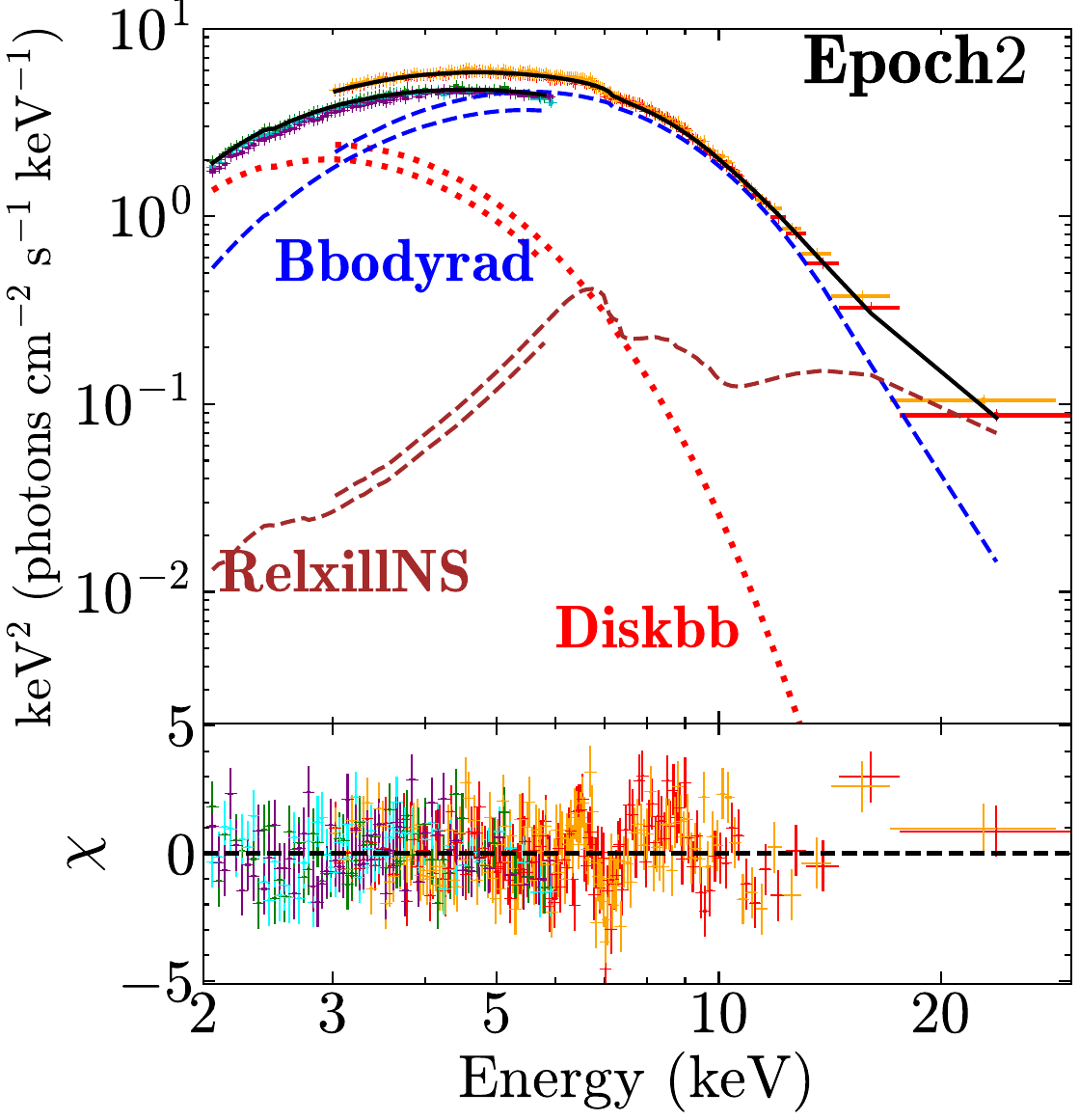}
\includegraphics[width=0.31\textwidth]{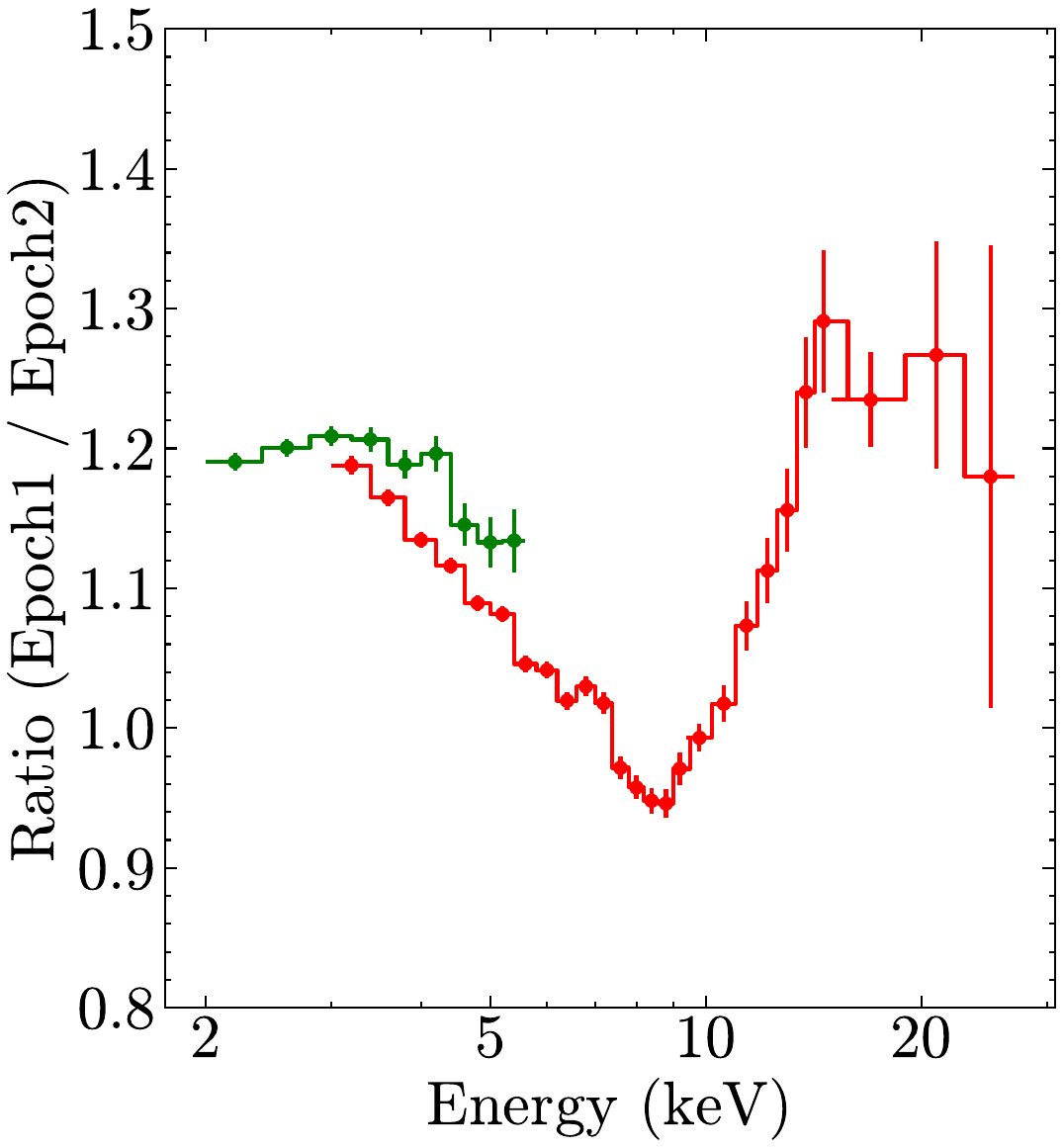}
\caption{Simultaneous \textit{NuSTAR} and \textit{IXPE} energy spectra with best-fit models in Epoch1 (left) and Epoch2 (middle) of Obs2. The right panel displays the spectral flux ratio between the two epochs, using \textit{IXPE} DU1 (green) and \textit{NuSTAR} FPMA (red) data.}
\label{fig:spectra}
\end{figure*}

The best-fit spectra with model components for the two epochs are shown in Figure~\ref{fig:spectra}, and the spectral parameters are listed in Table~\ref{tab:spectra_fitting}. 
We also plot the data flux ratio in the same figure, as a model-independent examination of spectral variation. 
The ratio spectrum exhibits two bumps, one in 2--9 keV and the other in 10--30 keV, corresponding to a higher disk temperature and higher reflection in Epoch1, as one can see from model parameters. The \texttt{Bbodyrad} component remains almost the same between the two epochs.

\begin{deluxetable}{llcc} 
\tabletypesize{\footnotesize}
\tablewidth{\columnwidth}
\tablecaption{Best-fit parameters for spectra in the two epochs.}
\label{tab:spectra_fitting}
\tablehead{
\colhead{Model}  & \colhead{Parameter} & \colhead{Epoch1} & \colhead{Epoch2}
}
\startdata
      Tbabs & $N_{\rm H}$ ($10^{22} \ \rm cm^{-2}$) & $3.65^{+0.19}_{-0.10}$ & $3.40^{+0.11}_{-0.15}$ \\
      Diskbb   & $T_{\rm in}$ (keV) & $0.93^{+0.11}_{-0.08}$ & $0.86\pm0.04$ \\
         & $R_{\rm in}$ (km) & $33^{+8}_{-6}$ & $33^{+5}_{-4}$ \\
      Bbodyrad   & $kT$ (keV) & $1.33^{+0.10}_{-0.04}$ & $1.35\pm0.01$ \\ 
         & $R_{\rm bb}$ (km) & $17^{+2}_{-3}$ & $17.2^{+0.7}_{-0.3}$  \\ 
         \noalign{\smallskip}\hline\noalign{\smallskip}
         RelxillNS   & Emissivity  & $3^\ast$ & $3^\ast$ \\
            & $R_{\rm in}$   (\(R_{\rm ISCO}\)) & $1^\ast$ & $1^\ast$ \\
            & $R_{\rm out}$   (\(R_{\rm g}\)) & $1000^\ast$ & $1000^\ast$ \\
            & $a$ & $0.1^\ast$ & $0.1^\ast$ \\
            & $\log N$ ($\rm cm^{-3}$) & $19^\ast$ & $19^\ast$ \\
            & Inclination (deg) & $31^{+2}_{-3}$ & $31^\ast$ \\
            & $\log_{\xi}$ & $2.64^{+0.12}_{-0.08}$ & $2.91^{+0.14}_{-0.10}$ \\
            & $A_{\rm Fe}$ & $3.0^{+0.9}_{-0.7}$ & $5.0^{+0.8}_{-0.7}$ \\
            & $kT_{\rm bb}$ (keV) & $2.69^{+0.10}_{-0.06}$ & $3.21^{+0.12}_{-0.10}$ \\
            & Norm ($10^{-3}$) & $2.75^{+0.25}_{-0.38}$ & $1.17^{+0.15}_{-0.13}$ \\ \noalign{\smallskip}\hline\noalign{\smallskip} 
        Cross-cal   & $K_{\rm FPMA}$  & 1$^\ast$ & 1$^\ast$ \\
         &  $\Delta \Gamma_{\rm FPMA}$ & 0$^\ast$ & 0$^\ast$   \\ 
            & $K_{\rm FPMB}$ & $1.01\pm0.01$ & $0.99\pm0.02$ \\
         &  $\Delta \Gamma_{\rm FPMB}$ ($10^{-3}$) & $3.5\pm6.3$ & $-13\pm9$   \\ 
            & $K_{\rm DU1}$ & $0.88\pm0.02$ & $0.91\pm0.03$ \\
         &  $\Delta \Gamma_{\rm DU1}$ ($10^{-3}$) & $17\pm20$ & $77^{+24}_{-22}$   \\ 
          & $K_{\rm DU2}$ & $0.84\pm0.02$ & $0.86\pm0.03$ \\
         &  $\Delta \Gamma_{\rm DU2}$ ($10^{-3}$)& $11^{+21}_{-20}$ & $52^{+25}_{-22}$  \\ 
          & $K_{\rm DU3}$ & $0.80\pm0.02$ & $0.75^{+0.03}_{-0.02}$ \\
         &  $\Delta \Gamma_{\rm DU3}$ ($10^{-3}$)& $10^{+21}_{-20}$ & $-29^{+24}_{-22}$  \\  
         \noalign{\smallskip}\hline\noalign{\smallskip}
         & $\chi^2 / {\rm d.o.f}$ & 1105.24/1096 & 1227.25/1001 \\ 
         \noalign{\smallskip}\hline\noalign{\smallskip} 
    Flux ratio    & $F_{\rm diskbb} / F_{\rm total}$ (\%)  &  44 & 32 \\ 
    (2--8 keV)    & $F_{\rm bbodyrad} / F_{\rm total}$  (\%)  &  51 & 65 \\
        & $F_{\rm relxillNS} / F_{\rm total}$  (\%)  &  5 &  3 \\
\enddata
\tablenotetext{^\ast}{Parameters fixed in the fit.}
\tablecomments{$R_{\rm in}$ and $R_{\rm bb}$ are derived assuming a distance of 10 kpc and an inclination of $0^{\circ}$. Please refer to \citet{Dauser2016} for the normalization of RelxillNS.}
\end{deluxetable}

\section{Discussions}
\label{sec:disc_conclu}

We re-analyzed the archival \textit{IXPE} observations of XTE J1701--462 during its 2022 outburst, and obtained time-averaged results well consistent with those reported in the literature \citep{Cocchi_xtej1701,Jayasurya_xtej1701,Yu_xtej1701}. However, we found that the nondetection of polarization in Obs2 was in fact due to a significant temporal variation of PA, in particular in Obs2-Epoch1, by an angle of $76^\circ \pm 8^\circ$ with respect to Obs1, $79^\circ \pm 11^\circ$ to Obs2-Epoch2, or $55^\circ \pm 11^\circ$ to Obs2-Epoch3 .

The PA variation is associated with spectral variation. As one can see in the CCD (Figure~\ref{fig:hid_ccd}), during Obs2-Epoch1, the source follows a track distinct from the tracks in Epoch2 and Epoch3, which are along the same trend. In the HID, the source count rate in Epoch1 is also among the highest in the NB. The spectral discrepancy is clearly identified in the energy spectrum (Figure~\ref{fig:spectra}). Compared with the source spectrum in Epoch2, the spectrum in Epoch1 shows excesses in both the low energy (peaked around 3--4~keV) and high energy (peaked around 20~keV) bands. The former is attributed to higher thermal disk emission, while the latter is due to an enhanced reflection component in Epoch1. Also, from the spectral ratio between the two epochs, there seems to be an excessive emission line component around 6--7 keV in Epoch1, but the significance is not high. 

In low magnetic systems, X-ray polarization usually reflects the geometry in radiative transfer, as the PA is perpendicular to the orientation of the last scattering direction. 
Thus, if the photons are scattered only once or the scattering mean free path is much larger than the corona size (case A), the relative location of the emission region and scattering region determines the PA. 
If the scattering mean free path is comparable to the corona size (case B), i.e., photons undergo several scatterings in the corona, the average PA is perpendicular to the elongation of the corona, because the last scattering has a higher chance to occur for photons traveling along the elongation \citep{ST85,Schnittman2010,Krawczynski2022,Tomaru_disc_wind}.
Otherwise, if the optical depth is extremely high or the scattering mean free path is much smaller than the corona size (case C), the PA is expected to be aligned with the elongation \citep[cf., the case of scattering plane-parallel atmosphere in][]{Chandrasekhar_1960}. We note that, a greater number of scattering is needed to boost photon energies into a higher energy band \citep{Rybicki_Lightman_1986}. Thus, even in an optically thin corona, the number of scattering could be greater than unity if the seed photon energy is much lower than the observing energy band (2--8 keV for \textit{IXPE}). Numerical simulations produce results in agreement with the first-principle estimation \citep[e.g.,][]{Schnittman2010}.

In the case of XTE J1701--462, emission from the accretion disk dominates the lower IXPE band while emission from the transition layer dominates the higher IXPE band. With spectro-polarimetry, \citet{Cocchi_xtej1701} revealed an energy dependent polarization and demonstrated that the polarization signal mainly arises from emission in the transition layer instead of the accretion disk in both Obs1 and Obs2. The data in Epoch1 alone do not allow us to perform a similar decomposition due to insufficient statistics. 
We assume that the spectral decomposition in Epoch1 is similar to that averaged in the whole Obs2, and the corona or transition layer is optically thick (modeled using a \texttt{Bbodyrad} model), approaching the case C.
Then, the PA variation may have revealed a transformation of the transition layer geometry, e.g., from a slab geometry to a spreading layer geometry or vice versa.  Assuming that the X-ray PA in NS-LMXBs during the horizon branch generally aligns with the jet orientation, as suggested by previous studies \citep{Farinelli_cygx-2, Bhargava_GX340+0}, the PA measured in Obs1 (horizontal branch) is expected to be perpendicular to the disk plane. The consistency of the PAs in Epochs 2 and 3 with that of Obs1 suggests that the corona geometry during these epochs is more vertically extended, resembling a spreading layer. In contrast, the PA variations observed in Epoch 1 may indicate a transition in the corona geometry, from a spreading layer to a slab-like configuration above the disk. Consequently, an enhanced reflection component is observed in Epoch 1 compared to Epoch 2.
Such a speculation can be confirmed if the radio jet in the source can be firmly detected in the future \citep{Fender_radio,Gasealahwe_etal_radio_2023}. 
We note that the observed PA variation could also be due to variation in the optical depth rather than geometry, or both. However, the spectral modeling seems not in favor of a remarkable change in the corona emitting spectrum (the \texttt{Bbodyrad} component).

\begin{acknowledgments}

We acknowledge funding support from the National Natural Science Foundation of China under grants Nos.\ 12025301, 12103027 \& 12122306, and the Strategic Priority Research Program of the Chinese Academy of Sciences.

\end{acknowledgments}

\facilities{IXPE, NuSTAR}

\bibliographystyle{aasjournal}
\bibliography{xtej1701}

\end{document}